\documentclass[sigconf]{acmart}

\usepackage{amsmath,amsfonts}
\usepackage{algorithm}
\usepackage{adjustbox}
\usepackage{lscape}
\usepackage{siunitx}
\usepackage[noend]{algpseudocode}
\usepackage{textcomp}
\usepackage{fancyvrb}
\usepackage{graphicx}
\usepackage{soul}
\usepackage{booktabs}

\usepackage{tikz}
\usetikzlibrary{arrows.meta, positioning}
\usepackage[utf8]{inputenc}
\usepackage[T1]{fontenc}
\usepackage{rotating}
\usepackage{graphicx}
\usepackage{paralist}
\usepackage{tabularx}
\usepackage{multicol}
\usepackage{multirow}
\usepackage{pbox}
\usepackage{enumitem}	
\usepackage{colortbl}
\usepackage{pifont}
\usepackage{xspace}
\usepackage{url}
\usepackage{tikz}
\usepackage{tabularx}
\usepackage{fontawesome}
\usepackage{lscape}
\usepackage{listings}
\usepackage{color}
\usepackage{showexpl}
\usepackage{anyfontsize}
\usepackage{comment}
\usepackage{soul}
\usepackage{multibib}
\usepackage{multirow}
\usepackage{todonotes}
\usepackage{xcolor}
\usepackage{xspace}
\usepackage{caption}
\usepackage{footnote}
\usepackage{tcolorbox}
\usepackage{booktabs}
\usepackage{fixltx2e}
\usepackage{balance}
\usepackage{ifthen}
\usepackage[normalem]{ulem}

\usepackage{lmodern}
\usepackage{microtype}
\usepackage{hyperref}
\usepackage{listings} 

\definecolor{codegreen}{rgb}{0,0.6,0}
\definecolor{codegray}{rgb}{0.5,0.5,0.5}
\definecolor{codepurple}{rgb}{0.58,0,0.82}
\definecolor{backcolour}{rgb}{1,1,1} 

\lstdefinestyle{pythonstyle}{
    language=Python,
    backgroundcolor=\color{backcolour},   
    commentstyle=\color{codegreen},
    keywordstyle=\color{magenta},
    numberstyle=\tiny\color{codegray},
    stringstyle=\color{codepurple},
    basicstyle=\ttfamily\footnotesize,
    breakatwhitespace=false,         
    breaklines=true,                 
    captionpos=b,                    
    keepspaces=true,                 
    numbers=left, 
    numbersep=5pt,                  
    showspaces=false,                
    showstringspaces=false,
    showtabs=false,                  
    tabsize=4,
    frame=none, 
    }

\AtBeginDocument{%
  }

\newboolean{showcomments}

\setboolean{showcomments}{true}

\ifthenelse{\boolean{showcomments}}
{\newcommand{\nb}[2]{
		\fbox{\bfseries\sffamily\scriptsize#1}
		{\sf\small$\blacktriangleright$\textit{#2}$\blacktriangleleft$}
	}
	
}
{\newcommand{\nb}[2]{}
	
}

\newcommand{\ie}{\emph{i.e.,}\xspace}
\newcommand{\eg}{\emph{e.g.,}\xspace}

\newcommand{\etal}{\emph{et~al.}\xspace}
\newcommand{\secref}[1]{Section~\ref{#1}\xspace}
\newcommand{\figref}[1]{Figure~\ref{#1}\xspace}

\newcommand{\tabref}[1]{Table~\ref{#1}\xspace}

\definecolor{lightergray}{rgb}{0.9,0.9,0.9}
\newtcolorbox{resultbox}{colback=lightergray, arc=0.5mm, top=2mm, bottom=2mm, left=2mm, right=2mm}

\newcommand{\tool}[1]{\texttt{\#}\xspace}
\newcommand{\rqone}{To what extent are MLOps frameworks used in open-source projects?}
\newcommand{\rqtwo}{What are the usage purposes of the MLOps frameworks in the studied projects?}
\newcommand{\rqthree}{What are the desired features or feature enhancements emerging from repository issues?}

\newcommand{\numFirstLevelInitial}{17\xspace}
\newcommand{\numFirstLevel}{14\xspace}
\newcommand{\numSecondLevelInitial}{54\xspace}
\newcommand{\numSecondLevel}{37\xspace}
\newcommand{\numIssues}{4,386\xspace}
\newcommand{\issueSampleSize}{311\xspace}
\newcommand{\notsupported}{\textcolor{red}{\ding{55}}}
\newcommand{\numRepo}{969\xspace}

\makeatletter 
\newcommand{\linebreakand}{%
\end{@IEEEauthorhalign}
\hfill\mbox{}\par
\mbox{}\hfill\begin{@IEEEauthorhalign}
}
\makeatother 

\copyrightyear{2026}
\acmYear{2026}
\setcopyright{cc}
\setcctype{by}
\acmConference[MSR '26]{23rd International Conference on Mining Software Repositories}{April 13--14, 2026}{Rio de Janeiro, Brazil}
\acmBooktitle{23rd International Conference on Mining Software Repositories (MSR '26), April 13--14, 2026, Rio de Janeiro, Brazil}
\acmDOI{10.1145/3793302.3793366}
\acmISBN{979-8-4007-2474-9/2026/04}

\begin{document}

\title[How are MLOps Frameworks Used in Open Source Projects?]{How are MLOps Frameworks Used in Open Source Projects? \\ An Empirical Characterization}

\author{Fiorella Zampetti}
\affiliation{%
 \institution{University of Sannio}\city{Benevento}\country{Italy}
}
\email{f.zampetti@unisannio.it}

\author{Federico Stocchetti}
\affiliation{%
 \institution{University of Sannio}\city{Benevento}\country{Italy}
}
\email{f.stocchetti1@studenti.unisannio.it}

\author{Federica Razzano}
\affiliation{%
 \institution{University of Sannio}\city{Benevento}\country{Italy}
}
\email{f.razzano2@studenti.unisannio.it}

\author{Damian Andrew Tamburri}
\affiliation{%
 \institution{University of Sannio}\city{Benevento}\country{Italy}
}
\email{datamburri@unisannio.it}

\author{Massimiliano Di Penta}
\affiliation{%
 \institution{University of Sannio}\city{Benevento}\country{Italy}
}
\email{dipenta@unisannio.it}

\renewcommand{\shortauthors}{Zampetti, Stocchetti, Razzano, Tamburri, and Di Penta}

\begin{abstract}
Machine Learning (ML) Operations (MLOps) frameworks have been conceived to support developers and AI engineers 
in managing the lifecycle of their ML models.
While such frameworks provide a wide range of features, developers may leverage only a subset of them, while missing some highly desired features.
This paper investigates the practical use and desired feature enhancements of eight popular open-source MLOps frameworks. Specifically, we analyze their usage by 
dependent projects on GitHub, examining how they invoke the frameworks' APIs and commands.
Then, we qualitatively analyze feature requests and enhancements mined from the frameworks' issue trackers, relating these desired improvements to the previously identified usage features.
Results indicate that MLOps frameworks are rarely used out-of-the-box and are infrequently integrated into GitHub Workflows, but rather, developers use their APIs to implement custom functionality in their projects. Used features concern core ML phases and whole infrastructure governance, sometimes leveraging multiple frameworks with complementary features.
The mapping with feature requests highlights that users mainly ask for enhancements to core features of the frameworks, but also better API exposure and CI/CD integration.
\end{abstract}



\maketitle

\section{Introduction}
\label{sec:intro}

Machine-Learning (ML) intensive applications are becoming extremely widespread in industry and open-source, also thanks to the increasing maturity of the field and the ease of entry-level ML-based software development. 

MLOps~\cite{john} has been put in place to properly manage the development process of such applications. 
Like DevOps for traditional software, MLOps provides a framework for managing ML-intensive systems' runtime, operations, and evolution. 
By enabling live monitoring, it supports making decisions such as triggering model retraining after having observed performance degradation \cite{{Bhaskhar2024,fedeli2023}}, or determining how much new data is needed to fine-tune a model~\cite{10185600}. 
Academic literature on the advantages of MLOps is still limited~\cite{calefato}, while industry data underscore the need for its adoption. A report from VentureBeat/IDC~\cite{columbus2023power} estimates that  80\% of analytics projects 
fail to move from pilot to production due to operational and reproducibility challenges MLOps could solve. This is confirmed by a Gartner report~\cite {gartnerInsights} that indicates an even higher percentage (87\%). 


In general, the practical use and understanding of MLOps in action seem limited \cite{calefato}, despite the availability of many open-source MLOps frameworks and the active communities around them.
To bridge this gap, this paper investigates \textbf{how MLOps frameworks are used in real-world open-source projects}, focusing on practical application and desired feature enhancements. Specifically, the study investigates (i) the extent to which different frameworks are used, (ii) the MLOps features being used, and (iii) the features requested in the frameworks' issue trackers.

 We select eight popular open-source MLOps frameworks, namely, \textsc{BentoML}~\cite{bentoml}, \textsc{Deepchecks}~\cite{deepchecks}, \textsc{Evidently AI}~\cite{evidently}, \textsc{Kedro}~\cite{kedro},  \textsc{Metaflow}~\cite{metaflow}, \textsc{MLFlow}~\cite{mlflow}, \textsc{Prefect}~\cite{prefect}, and \textsc{Wandb}~\cite{wandb}. The selection includes frameworks covering the three key MLOps components: ML workflow management, model deployment, and production monitoring.
By analyzing \numRepo Python GitHub repositories using the eight frameworks (obtained from GitHub dependents after a proper filtering), we study the usage mode, \eg, as APIs in Python scripts or from GitHub workflows, and categorize the
MLOps features used. Then, to understand users' needs, we first manually categorize a statistically significant sample of \issueSampleSize feature request issues, and then complement this analysis automatically with a Large Language Model (LLM), covering the total set of \numIssues feature requests for the eight frameworks.

Our findings highlight that MLOps frameworks are rarely used out-of-the-box---\eg simply running them from the command line---or in GitHub Workflows. Instead, developers leverage the frameworks' APIs to build customized MLOps workflows. In terms of used features, developers leverage them for model training, monitoring, and infrastructure governance. Developers combine multiple frameworks mainly because of the different features they provide, thereby composing them in a pattern-based fashion that deserves further investigation.
The analysis of frameworks' feature requests led to a taxonomy of \numFirstLevel categories, organized into \numSecondLevel sub-categories, covering different phases of the ML pipeline, or related to its governance. 
Last but not least, the mapping between used features and desired enhancements highlights that while users often ask to improve core MLOps features, they also ask for changes to their API/interfaces---given that this is their primary way of use. Also, better integration with CI/CD pipelines is often requested, along with improvements in monitoring, to keep track of the pipeline's health. 

The contributions of this paper can be summarized as:
\begin{compactitem}
\item An analysis of features of eight popular open-source MLOps frameworks used by \numRepo clients;
\item A taxonomy of feature requests for the eight frameworks, and how they relate to their usages; and
\item A replication package featuring datasets and scripts used in our analyses \cite{replication}.
\end{compactitem}

\section{MLOps Frameworks: Background}

MLOps is an emerging practice that adapts DevOps principles for end-to-end ML pipelines. The goal of MLOps is to enable teams to efficiently manage data, train models at scale, and ensure ``reliable'' deployment for predictive applications~\cite{john}.
This reliability is achieved by combining automation, such as artifact versioning and reproducible training, with production monitoring. The latter creates a feedback loop, ensuring models perform stably and adapt as data and user needs change. 


Graph-based automation frameworks exist~\cite{JohnOB25,warnett}, helping manage the end-to-end ML lifecycle. For instance, \textsc{MLFlow} is an open-source framework that provides features for experiment tracking, model packaging, and model registry. As shown in Listing \ref{lst:mlflow}, \textsc{MLFlow} allows the creation of experiments within which each run records parameters, metrics, tags, and artifacts (\eg plots or serialized models), enabling transparent comparison of runs and easy reproduction. Furthermore, it has a Dashboard supporting quick and easy inspection of results, \eg \texttt{mlflow ui}. Furthermore, \textsc{MLFlow} provides a framework-agnostic format for packaging and deploying models, a Model Registry for versioning artifacts, and autologging for automatic tracking. Those features enhance reproducibility and collaboration.

While frameworks exist to support the development of ML-intensive systems, the MLOps ecosystem continues to evolve rapidly, attracting the research community, which started to define best practices emphasizing robust governance, explainability, compliance, and better alignment between ML teams and domain experts. We conclude that MLOps as a discipline is poised to become essential for any organization sustaining ML applications at scale.

\begin{lstlisting}[style=pythonstyle, caption={Hello \textsc{MLFlow}: log a param, metrics, and a sample artifact.}, label={lst:mlflow}]
# hello_mlflow.py
from pathlib import Path
import time

mlflow.set_tracking_uri("file:./mlruns") # local store

# set the experiment
mlflow.set_experiment("hello-mlflow") 

with mlflow.start_run(run_name="first-run") as run:
    # Log hyperparameters
    mlflow.log_param("learning_rate", 0.1)
    
    # Log the loss metric history
    for step, loss in enumerate([1.0, 0.7, 0.5, 0.4, 0.35], start=0):
        mlflow.log_metric("loss", loss, step=step)

    # Log an artefact to make it visible in the Dashboard
    notes = Path("notes.txt")
    notes.write_text("Hello, MLflow! This is an artifact.")
    mlflow.log_artifact(str(notes))

print("Done. View results with `mlflow ui` and open http://127.0.0.1:5000")
\end{lstlisting}

\section{Empirical Study}
\label{sec:study}

The \emph{goal} of the study is to investigate how MLOps frameworks are used in open-source projects. The \emph{quality focus} is on understanding the usage modes and features of MLOps frameworks in ML-intensive applications. 
The \emph{perspective} is of researchers interested in supporting software developers in evolving their ML-intensive applications through MLOps frameworks, by identifying typical usages and evolution (through feature requests) patterns. 
The \emph{context} consists of \numRepo open-source Python projects hosted on GitHub, importing at least one of the eight MLOps frameworks, objects of the study. 
Based on the stated goal, our study aims to address the following three research questions (RQs):

\begin{itemize}
    \item  \textbf{RQ$_1$:} \emph{\rqone} In this preliminary research question, we look at the extent to which MLOps frameworks are used in open-source projects as APIs in the production code, as well as in the GitHub Continuous Integration and Delivery (CI/CD) process, \ie in steps executing them in the GitHub Workflow pipeline. 
    \item \textbf{RQ$_2$:} \emph{\rqtwo}  
     We identify MLOps framework API usages and map them to the phases of an ML workflow, such as data preparation, feature engineering, model training, model validation, model deployment, ML workflow governance, and monitoring. Furthermore, 
    given that frameworks may offer overlapping or complementary features, we also analyze the co-usage patterns in projects that rely on more than one framework. 
    \item \textbf{RQ$_3$:} \emph{\rqthree} Besides understanding how MLOps frameworks are being used, we are
    interested in figuring out whether such frameworks exhibit needs for improvements---in terms of completely new features or enhancement of existing ones---as it emerges by looking and classifying fixed repository issues. 
    \end{itemize}

\begin{figure}[t]
    \centering
    \includegraphics[width=1\linewidth]{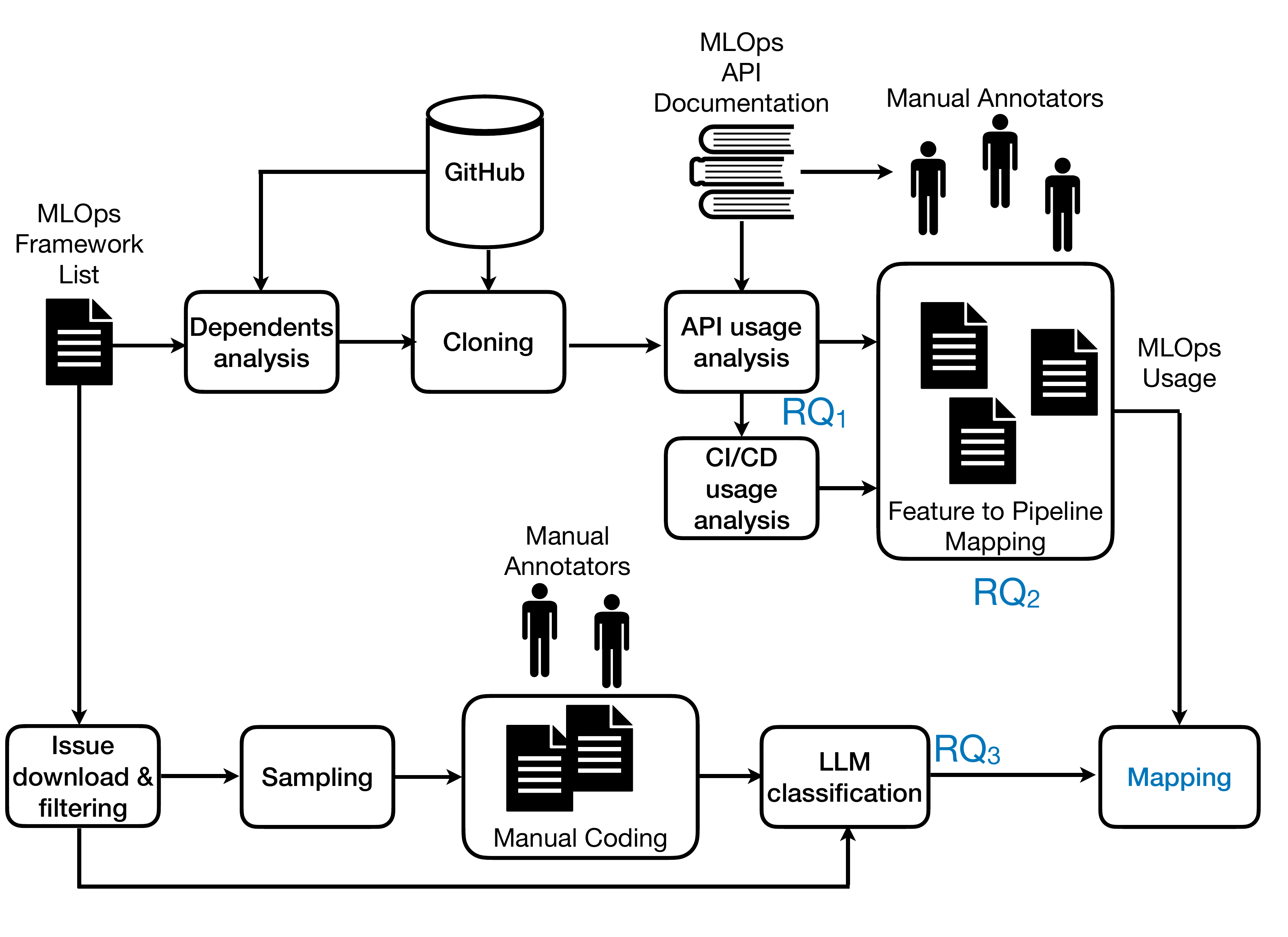}
    \caption{Data analysis process.}
    \label{fig:process}
\end{figure}

\figref{fig:process} overviews our analysis process. We first identify projects based on MLOps frameworks to analyze their usage (RQ$_1$ and RQ$_2$). Then, we combine manual analysis and LLMs to categorize frameworks' feature requests (RQ$_3$). As a last step, we relate the framework usages with the community's requested features.

\subsection{MLOps framework Selection}
\label{sec:mlopstools}

To select relevant MLOps frameworks, due to the relative infancy of the phenomenon under investigation, we analyzed blogs \cite{Awan2024,Kelvins2025} and framework lists \cite{recupito}, providing an overview of the MLOps landscape. The selection process focused on open-source frameworks hosted on GitHub, making it easy to identify projects that use them (see \secref{sec:dataset}). Furthermore, we focused the analysis on frameworks for ML applications, excluding frameworks focusing on very specific LLM aspects, e.g., LLM testing.
Note that, we included only the frameworks aimed at (i) managing model metadata and supporting experiment tracking, \ie \textsc{MLFlow} and \textsc{Wandb}, (ii) managing the whole ML workflow, \ie  \textsc{Kedro}, \textsc{Metaflow}, and \textsc{Prefect}, (iii) testing the data/model quality also in production, \ie \textsc{Deepchecks} and \textsc{Evidently AI}, and (iv) deploying ML models, \ie \textsc{BentoML}.
As a final selection criterion, we only included frameworks that support Python and common ML libraries such as TensorFlow, PyTorch, and Keras.
\tabref{tab:tools} provides an overview of the eight MLOps frameworks selected for the study, including the number of dependents and issues.

\begin{table}[t]
	\caption{MLOps frameworks considered in the study}  
	\label{tab:tools}
	\centering
	\small
    \resizebox{\columnwidth}{!}{
	\begin{tabular}{l|rr|rr}
		\hline
        \hline
        \textbf{Name} & \textbf{\# Dep.} & \textbf{\# Proj.} & \textbf{\# Issues} & \textbf{\# Feat. Req.}  \\ 
		\hline
        \hline
        \textsc{BentoML~\cite{bentoml}} & 
        1,747 & 31 & 1,102 & 94\\
        \textsc{Deepchecks~\cite{deepchecks}} & 
        364 & 9 & 979 & 404\\
        \textsc{Evidently AI~\cite{evidently}} & 
        3,936 &33 & 428 & 67 \\
        \textsc{Kedro~\cite{kedro}} & 
        1,917 & 21 & 2,149 & 618\\
        \textsc{Metaflow~\cite{metaflow}} & 
        520 & 26 & 689 & 67 \\
        \textsc{MLFlow~\cite{mlflow}} & 
        35,297 & 644 & 4,120 & 1,024 \\
        \textsc{Prefect~\cite{prefect}} & 	
        4,452 & 105 & 6,051 & 1,598\\
        \textsc{Wandb~\cite{wandb}} & 
        5,692 & 688 & 3,516 & 514 \\
        \hline
        \hline
	\end{tabular}}
    \vspace{-3mm}
\end{table}

\subsection{Dataset Collection}
\label{sec:dataset}
After selecting MLOps frameworks, we identify GitHub projects using them by scraping---using \texttt{Beautiful Soup}~\cite{beautifulsoup:2023} and \texttt{requests} Python packages---
their list of dependents from the GitHub dependency graph.
The second column of \tabref{tab:tools} reports the total number of dependents collected for each framework. 
We filtered out the projects with fewer than 10 stars to avoid toy projects. 
Also, through regular expressions on repository names and descriptions, if available (see the replication package), we filtered out 
tutorials, code books, and example projects.
Furthermore, we retained only Python projects, since it is the ``de facto'' language for developing ML applications. 
The third column of \tabref{tab:tools} reports, for each framework, the total number of repositories after applying our filtering procedure. The table shows that the number of repositories resulting from our filtering is a small fraction of the initial set, \ie only 1,557 out of 53,925 projects (2.88\%) met our criteria. This suggests how, currently, MLOps frameworks are predominantly used within tutorials, personal examples, and small study/research projects.

\subsection{Usage Extraction}
To address RQ$_1$, for each framework, we search for its usage in (a) production code---with a focus on Python, the \emph{lingua franca} for ML applications---and (b) GitHub workflows. After cloning each repository, we look at Python files that import the framework by analyzing the imports through the Python AST package. Also, as data scientists mainly rely on Jupyter Notebooks~\cite{zhang2020data, zou2024can}, we transform each ``.ipynb'' into a Python file by using the \texttt{nbconvert} tool~\cite{nbconvert}. After the conversion, we use a similar procedure to check whether the notebook imports the framework. 

Upon identifying the source code files, we extract all framework usages in each file by using the Python AST package, searching for:
\begin{compactitem}
    \item Function calls referencing the MLOps framework, \eg \texttt{bentoml.picklable\_model.load\_runner(...)}. If a returned value is assigned to a variable, we also track every statement where that variable is used.
    \item Object instantiations with their assignment to variables, \eg \texttt{flow = Flow(...)}. Just like before, we also trace where this variable is used.
    \item Attribute accesses for the objects instantiated relying on classes available in the MLOps framework API, \eg \texttt{latest = flow.latest\_successful\_run}.
    \item Definitions of new classes extending classes inherited from the MLOps framework API, \eg \texttt{def class RegressionModel(FlowSpec)}. In this case, we also store the set of functions (methods) defined as part of the new class definition.
    \item For \textsc{Metaflow} and \textsc{Prefect}, as their usage model consists of instantiating a pipeline object, we also extract the signature of the functions attached to decorators defining steps/tasks used to create and define an ML pipeline. Specifically, for \textsc{Metaflow}, we consider the \texttt{@step} decorator, while for \textsc{Prefect} we consider \texttt{@task} and \texttt{@flow} decorators. 
\end{compactitem}

Concerning framework usages in CI/CD pipelines, we focus on GitHub workflows, being, to date, by far the most popular CI/CD tool used on GitHub \cite{MazraeMGD23}. We first check whether the project relies on GitHub workflows by searching for ``.github/workflows/'' folders. Then, we analyze the found ``.yaml'' workflows using the Python \texttt{yaml} package to extract the set of defined jobs, along with any steps specifying a \texttt{run} command. However, given that not all MLOps frameworks have a command-line interface, we defined two different regular expression-based search strategies:
    (i) Check whether the framework is used from the command line interface (CLI), \eg \texttt{run: bentoml build}; and 
    (ii) Check whether any of the production files importing the MLOps framework, as identified in the previous command line analysis, is executed directly as part of the \emph{run:\$} command, \eg \texttt{run: python application/src/save\_model\_to\_bentoml.py}~where \texttt{save\_model\_to\_bentoml.py} imports \textsc{bentoML}.
As a result, for each client project that depends on an MLOps framework, we identify source code files that utilize it, the specific statements in which it is invoked, and the workflows that directly or indirectly use the framework.

\subsection{Usage Classification}
To answer RQ$_2$, the authors assigned a list of core features to each framework by looking at their official documentation. After that, from the statements using the MLOps framework API, we extracted the API being invoked. By relying on the official documentation of each framework, the first author manually assigned a specific step of the ML development and monitoring workflow, also accounting for the governance of the ML workflow pipeline. Note that RQ$_2$ does not aim to create a usage taxonomy through open coding; instead, we categorized the frameworks’ APIs along the ML phases as described by Amershi \etal~\cite{AmershiBBDGKNN019}, while also accounting for the core features provided by the analyzed frameworks. In case of doubt, the first author discussed the step with all other authors before assigning it. During this manual analysis, all low-level API usages that did not belong to the framework's core features were marked as ``Irrelevant''. For example, for \textsc{Prefect}, we found projects using the \texttt{prefect.utilities.graphql.compress()}
method, a low-level utility function used to construct GraphQL queries, \ie this is a utility feature supporting a core one.

\subsection{Feature Request Analysis}
To address RQ$_3$, we first use GitHub APIs to mine all closed and fixed issues across the eight MLOps frameworks (the fourth column of \tabref{tab:tools}).
Subsequently, only issues that are labeled as ``feature request", ``feature", ``enhancement" or, if these labels are not used, matching ``feature" in the title, are retained (fifth column of \tabref{tab:tools}),  leaving us with a total of \numIssues issues. The focus on closed, fixed issues ensures that our analysis focuses on features requested that were actually addressed by the maintainers.
After that, we extract a statistically significant sample using random stratified sampling, where strata are represented by frameworks. The sample size is 311, which allowed us to achieve 95\% significance with $\pm 5\%$ confidence interval.
We perform an annotation using a top-down approach because our primary goal is to map the feature requests on the MLOps pipeline and relate them to the feature usages studied in RQ$_2$, and only after delving into specific sub-features.

First, two independent (human) annotators performed an initial open coding of issues along broad topics. The annotation has been performed using the 
AtlasTI tool \cite{atlasti2024}. 
In the end, they obtained \numFirstLevelInitial categories, achieving, during the coding phase, a Cohen's $k$ inter-rater agreement~\cite{hsu2003interrater} of 62\% (moderate/substantial agreement). 
Then, a second classification was performed to identify lower-level feature categories and understand what needs to be improved in the MLOps frameworks.   To this aim, a human annotator performed a first categorization, leading to the identification of \numSecondLevelInitial sub-categories. 
Then, we leveraged an LLM---as also suggested in previous work by Ahmed \etal~\cite{AhmedDTP25}---to perform a second annotation into sub-categories, for which a mapping to high-level categories already existed. After trying five different LLMs, \ie Gemini 2.5 Flash, Gemini 2.5 Pro, Deepseek r1, Gpt-4o, and o3, and manually checking a subset of results (between 20 and 50 random instances) for each, we ended up selecting Gemini 2.5 Pro~\cite{google2025gemini2.5pro}. 

The LLM-based annotation was performed by asking it to classify each issue (from its title and body) into the human-defined categories and to highlight cases for which no suitable category existed (final version of the prompt available in our replication package). The LLM achieved a Cohen's $k$~\cite{hsu2003interrater} of 0.78, indicating strong agreement. A second author discussed with the manual annotator to resolve the discrepancies with the LLM in the manual classification. 
After that, the LLM was asked to classify the remaining \numIssues-311=4,075 issues. The LLM indicated that 153 did not belong to any category. Of these, 82 were classified as not being feature requests. This leaves us with only 153-82=71 (1.6\% of the total) issues uncovered by the taxonomy, which suggests a satisfactorily-enough level of saturation. 

After that, two further authors reviewed the produced taxonomy and consolidated both categories and sub-categories, performing some merges. This resulted in a final taxonomy of \numFirstLevel categories and \numSecondLevel sub-categories. Finally, the taxonomy has been further validated by one author who is an expert in the field.

\begin{table}[t]
	\caption{Overview of the MLOps Frameworks' Usages}  
	\label{tab:toolsUsage}
	\centering
	\small
	\begin{tabular}{lrrr}
		\hline
        \hline
		\textbf{Framework} & \textbf{\# Proj.} & \textbf{\# Proj. Using}  & \textbf{CI/CD use}\\ 
		\hline
        \hline
        \textsc{BentoML} & 31 & 22 & 1\\
        \textsc{Deepchecks} & 9 & 5 & 0 \\
        \textsc{Evidently AI} & 33 & 23 & 1 \\
        \textsc{Kedro} & 21 & 15 & 2 \\
        \textsc{Metaflow} & 26 & 18 & 0\\
        \textsc{MLFlow} & 644 & 400 & 8 \\
        \textsc{Prefect} & 105 & 79 & 10 \\
        \textsc{Wandb} & 688 & 447 & 2\\
        \hline
        \hline
	\end{tabular}
\end{table}

\section{Study Results}
\label{sec:results}
In this section, we present and discuss the results of each research question defined. 

\subsection{\textbf{RQ$_1$:} MLOps Frameworks Usage}

\tabref{tab:toolsUsage} reports, for each MLOps framework, the number of client projects among those we filtered that use the framework. As shown, 
\textsc{Wandb} and \textsc{MLFlow} are the most used frameworks. This is not surprising since they mainly provide features for model tracking, \eg Model Training, and for governing the whole ML pipeline. Instead, frameworks aimed at supporting developers in creating reproducible, maintainable, and modular ML pipelines, such as \textsc{Prefect}, \textsc{Metaflow}, and \textsc{Kedro}, are used by only a few projects. As evident from the frameworks' own documentation, the integration of such frameworks may require an initial not negligible effort to adapt one's own code to a specific paradigm, which in turn requires understanding and defining flows as a set of interconnected steps with dependencies, whereas the usage of tools like MLFlow requires the addition of a few code lines, \eg \texttt{mlflow.start\_run()} followed by \texttt{mlflow.log\_metric()}.  
Frameworks like \textsc{Deepchecks} and \textsc{Evidently AI}, which support developers in defining checks for data and model validation and monitoring, are very rarely used, with 5 and 23 projects, respectively. Since quality gates for monitoring performance degradation or drift after model deployment are mainly required at the production stage, this may indicate the lack of maturity of ML-intensive open-source projects, which seldom require MLOps automation for this phase.

As regards the use of MLOps frameworks in CI/CD pipelines (fourth column \tabref{tab:toolsUsage}), we found 24 projects that use an MLOps framework as a step in the GitHub Workflow, accounting for 35 ``.yml" files and 68 steps. Among these, 43 steps execute a pre-defined workflow/process managed by the framework's execution engine, such as \texttt{kedro run} or \texttt{prefect deployment build}. In comparison, 25 steps run a script executing a Python file that imports the MLOps framework. This result is somewhat surprising. While not all MLOps frameworks have CLIs suitable for direct invocation in GitHub Workflows, even those primarily used as libraries are rarely used in CI/CD pipelines. This points out a gap in MLOps maturity, where basic automation exists. Nevertheless, the adoption of dedicated frameworks for granular experiment tracking, data validation, and monitoring within those automated pipelines is not widespread. 

\subsection{\textbf{RQ$_2$:} Frameworks Features Being Used}

\begin{table*}[t]
    \caption{How MLOps frameworks are used}
    \label{tab:tools_usages}
    \centering
    \scriptsize 
    \setlength{\tabcolsep}{4pt} 
    \begin{tabular*}{\textwidth}{@{\extracolsep{\fill}}l r r r r r r r r r r}
    \toprule
        \multirow{2}{*}{\textbf{Framework}} & \multicolumn{1}{c}{\textbf{Data}} & \multicolumn{5}{c}{\textbf{ML Workflow Process}} & \multicolumn{2}{c}{\textbf{Model Lifecycle}} & \multicolumn{2}{c}{\textbf{Operations}} \\
        \cmidrule(lr){2-2} \cmidrule(lr){3-7} \cmidrule(lr){8-9} \cmidrule(lr){10-11} 
         & \textbf{Integrity} & \textbf{Def.} & \textbf{Load} & \textbf{Save} & \textbf{Run} & \textbf{Gov.} & \textbf{Train.} & \textbf{Valid.} & \textbf{Deploy} & \textbf{Monitor} \\
    \midrule \midrule 
    \textsc{BentoML}    & \notsupported & \notsupported  & 8   & 8   & 11  & 10 & \notsupported   & \notsupported  & 21 & 5   \\
    \textsc{Deepchecks}   & 4 & \notsupported  & \notsupported   & \notsupported   & \notsupported   & \notsupported  & \notsupported   & 4  & \notsupported  & 0   \\
    \textsc{Evidently AI}  & 14 & \notsupported  & \notsupported   & \notsupported   & \notsupported   & \notsupported  & \notsupported   & 19  & \notsupported  & 21   \\
    \textsc{Kedro}        & \notsupported & 15 & 9   & 0   & 12  & 13 &  \notsupported  & \notsupported   & \notsupported   & \notsupported   \\ 
    \textsc{Metaflow}      & \notsupported  & 18 & 4   & 0   & 0   & 8  & \notsupported   & \notsupported   & \notsupported  & 9 \\
    \textsc{MlFlow}       & \notsupported  & \notsupported  & 30 & 15 & 114 & 278& 189 & 9  & 5  & 173 \\ 
    \textsc{Prefect}      & \notsupported & 77 & 0   & 0   & 9   & 39 & \notsupported  & \notsupported  & 22 & 11  \\ 
    \textsc{Wandb}  & \notsupported & \notsupported & 4   & 40  & 31 & 188& 388 & 20 & \notsupported  & 406 \\ 
    \bottomrule
    \end{tabular*}
\end{table*}

To answer RQ$_2$, 
we counted the number of projects using the features made available by the frameworks. Results are summarized in \tabref{tab:tools_usages}, where a ``\notsupported'' symbol indicates that the framework does not support a given feature, whereas a ``0'' indicates that the feature is supported but not used by any project in our dataset. 
As the table shows, almost all key features are used by at least one project for each framework. The only exception is \textsc{Deepchecks}, where the Monitoring feature was never used. However, note that \textsc{Deepchecks} was found to be used only by five projects; therefore, this indication may not be very representative.

\textsc{BentoML} is mainly used to deploy a trained ML model in a production environment (21 out of 22 projects). This is achieved by instantiating the \texttt{BentoService()} class or the \texttt{Service()} API call to bundle trained models and inference logic into a container to be used for deployment. In doing this, developers can use specific API calls to properly set the infrastructure in terms of specifying the artifacts to be included (\texttt{@artifacts}), setting the runtime environment (\texttt{@env}), and defining the exposed APIs for inference by using the \texttt{@api} decorator. Once defined, developers rely on \texttt{bentos.build()} API call to concretely build a Bento, \ie packaged version of a \textsc{BentoML} service ready for deployment, or the \texttt{bentos.build\_bentofile()} call to create a Bento package starting from a configuration file. Furthermore, 11 projects use \textsc{BentoML}'s Runners to efficiently handle inference. \textsc{BentoML} can be invoked from the command line. We found only one project\footnote{\url{https://github.com/khuyentran1401/employee-future-prediction}} with a GitHub workflow (\ie deploy\_app.yaml) that trains the model and saves it to the \textsc{BentoML}, and, by executing the command \texttt{bentoml build}, builds the Bento package. The same project also includes a workflow for testing the model against the \textsc{BentoML} service locally, which starts with the command \texttt{bentoml serve}.

\textsc{Deepchecks} and \textsc{Evidently AI} support the validation and monitoring of ML models and data, to detect issues such as drift, data integrity problems, and performance degradation. 
Unlike \textsc{Deepchecks}, \textsc{Evidently AI} is mainly used for live monitoring, \ie 21 out of 23 projects. To achieve this, projects mainly call the \texttt{ModelMonitoring()} API to set up automated monitoring of ML models in production by providing as parameters monitors like \texttt{RegressionPerformanceMonitor} and  \texttt{ClassificationPerformanceMonitor} to detect, as soon as possible, when model predictions start becoming less accurate. As regards data integrity, developers rely on API calls such as \texttt{ConflictPredictionMetric()} and \texttt{ConflictTargetMetric()} to detect conflicts between predictions and target values in a dataset or inconsistencies in the target values. 
Regarding framework use in GitHub Workflows, we only found one project\footnote{\url{https://github.com/Alaboy19/model-retraining-gitops-fastapi/}} running a script (\ie \texttt{train.py}) importing and using \textsc{Evidently AI} to check for the presence of data drift across features, \ie \texttt{DataDriftTestPreset()}.

\textsc{MLFlow} and \textsc{Wandb} provide features to support model experiment tracking, \ie Model Training and Model Validation.  As reported in \tabref{tab:tools_usages}, we found 189 and 388 projects using each framework for training, while only 9 and 20 for Model Validation. Furthermore, \textsc{MLFlow} and \textsc{Wandb} are widely used for setting up and managing the entire ML infrastructure. For example, \textsc{campusx-official/yt-comment-sentiment-analysis} has eight Python files that import and use \textsc{MLFlow}. It instantiates the \texttt{MlflowClient()} to interact with both the MLflow Tracking Server and Model Registry, and uses it to run a set of experiments, properly set up using \texttt{set\_experiment()} and similar API calls, for model training and validation whose results are tracked by invoking API calls such as \texttt{log\_param()} and \texttt{log\_metric()}. Then, the model is added to the Model Registry via \texttt{model\_register()}. Model promotion is automated in the \textit{promote\_model.py} script, using the \texttt{client.transition\_model\_version\_stage()} to move a specific version to the ``Production'' stage. This script is executed in the GitHub Workflow within the ``model-deployment'' job under the step \textit{``Promote model to production''}. This example shows how \textsc{MLFlow} can be used for testing and evaluating different models and, after choosing the most suitable one, for deployment in production environments. The \textsc{MLFlow} deployment feature is used only 5 out of 400 projects.
It is important to note that the \textsc{Wandb} monitoring feature is used by most projects (406/447) that use the framework. However, note that such a monitoring feature primarily serves to track experiments rather than provide live monitoring in a production environment. 
This is done by instantiating objects like \texttt{Histogram()} to visualize how weights, activations, gradients, or outputs evolve during training or evaluation, or by invoking the \texttt{log()} or the \texttt{define\_metric()}. 

\begin{table}[t]
	\caption{Steps defined when creating an ML pipeline}  
    \label{tab:usages_pipelines}
	\centering
	\small
	\begin{tabular}{lrrr}
    \hline \hline
    \textbf{Step} & \textsc{Evidently AI} & \textsc{Kedro} & \textsc{Prefect}\\
    \hline \hline
    Data Collection & 4 & 3 & 38 \\
    Data Processing & 12 & 12 & 43 \\
    Data Labeling & 2 & 0 & 0 \\
    Data Integrity Check & 4 & 0 & 6 \\
    Feature Engineering & 5 & 3 & 9 \\
    Model Training & 10 & 10 & 21 \\
    Model Validation & 6 & 7 & 11 \\
    Deployment & 2 & 0 & 14 \\
    Model Inference & 3 & 0 & 8 \\
    Monitoring & 1 & 0 & 5 \\
    \hline \hline
    \end{tabular}
\end{table}

\begin{table}[t]
	\caption{MLOps frameworks co-usages}  
    \label{tab:usages_patterns}
	\centering
	\small
    \begin{tabular}{lr}
    \hline
    \hline
    \textbf{Pattern} & \textbf{\#Pr.} \\
    \hline \hline
    MLFlow, Prefect & 8 \\
    MLFlow, Evidently AI & 6 \\
    MLFlow, Kedro & 3 \\
    MLFlow, BentoML & 2 \\
    MLFlow, Metaflow & 2 \\
    MLFlow, Evidently AI, Prefect & 2 \\
    MLFlow, Deepchecks, Evidently AI & 1 \\
    MLFlow, Kedro, Prefect & 1 \\ 
    MLFlow, BentoML, Evidently AI, Prefect & 1 \\
    MLFlow, Deepchecks, Evidently AI, Prefect & 1 \\ 
    MLFlow, Deepchecks, Evidently AI, Metaflow, Prefect & 1 \\
    Kedro, Prefect & 1 \\
    \hline \hline 
    \end{tabular}
\end{table}

\textsc{Kedro}, \textsc{Metaflow}, and \textsc{Prefect} are mainly used to set up and define an ML pipeline. Each uses a specific paradigm to define nodes representing the ML workflow phases and their interconnection.
They also support properly orchestrating and governing the pipeline's use when setting up experiments and running them. As an example, \textsc{DAGWorks-Inc/hamilton} defines two pipelines, each comprising three phases. The first pipeline aims to preprocess the data and organize them to make them suitable for training and evaluation, \ie \textit{preprocess\_companies, preprocess\_shuttles, create\_model\_input\_table}. Instead, the second pipeline is used to train and evaluate the model using three phases: \textit{split\_data, train\_model, evaluate\_model}. The defined pipelines are subsequently executed by instantiating a runner (\texttt{SequentialRunner()}), and invoking the \texttt{run()} method to execute the workflow/pipeline specified as an argument. 
Note that the Monitoring feature in these MLOps frameworks is not intended for model monitoring but rather for checking the pipeline's health.  For this purpose, \textsc{Metaflow} provides the \texttt{@card} decorator that allows pipeline steps to automatically generate reports or plots. As an example, in the project \textsc{jacopotagliabue/post-modern-stack}, two steps are defined with the @card decorator, namely \textit{run\_transformation} to apply a transformation to the raw input data, and \textit{test\_model} to load a trained model and test its performance. This also applies to \textsc{Prefect}, where developers can use (i) the classes in the \texttt{states} module to store the status of each step/workflow defined for a specific experiment or (ii) rely on the \texttt{get\_run\_logger()} to specify different severity levels about the information to be logged and displayed durig the run of a step. Since \textsc{Prefect} also supports deploying trained and evaluated models, the \texttt{error()} severity level (which is pretty high) is rarely used and mainly applies to steps that set the deployment environment or deploy the model in production. 

\begin{figure*}[t]
    \centering
    \includegraphics[width=1\linewidth]{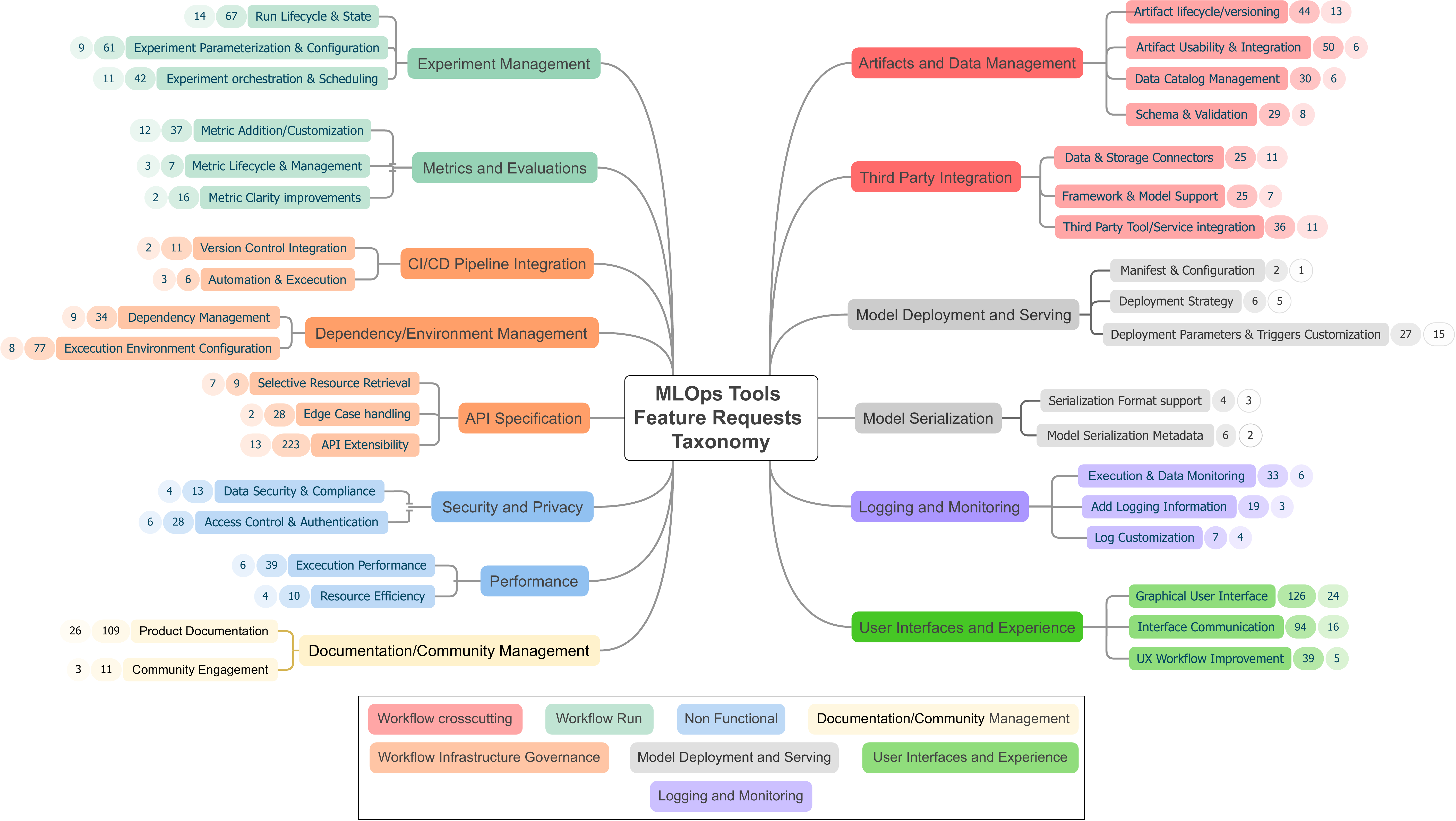}
    \caption{Taxonomy of feature requests from MLOps frameworks issue trackers.}
    \label{fig:taxonomy}
\end{figure*}

For the three MLOps frameworks that support ML pipeline definition (\textsc{Metaflow}, \textsc{Kedro}, and \textsc{Prefect}), we manually analyzed the created pipelines and mapped their steps/phases to the ML development phases. Results of this analysis are summarized in \tabref{tab:usages_pipelines}, where, for each pipeline step, we report the number of projects that integrate it when using a given framework to set up the ML pipeline. Unsurprisingly, the most common steps involve data collection, model training, and model validation. As an example of data processing and feature engineering, the project \textsc{v6d-io/v6d} defines a pipeline made up of five steps, all dealing with data cleaning before training\footnote{\url{https://github.com/v6d-io/v6d/blob/main/python/vineyard/contrib/kedro/benchmark/mlops/src/mlops/pipeline.py}}: removing outlier values, extracting the target variable, removing unnecessary columns from the feature set, handling missing data in according to a specified strategy, and creating a new feature by combing other features. As regards data labeling, we found two projects in our dataset where the pipeline defines a step requiring the connection to the \texttt{Argilla} workspace\footnote{\url{https://docs.argilla.io/latest/}}, \ie an annotation platform, to get annotated data to be used as input to train a model\footnote{\url{https://github.com/openfoodfacts/openfoodfacts-ai/blob/develop/spellcheck/scripts/dags/extract_from_argilla.py}}. Last but not least, we discuss one example\footnote{\url{https://github.com/IzicTemi/e2e_fake_news_classifier/blob/main/prefect_monitoring.py}} in which one of the steps defined in the pipeline by using \textsc{Prefect} leverages a monitoring through \textsc{Evidently AI} and \textsc{MLFlow}. Specifically, the pipeline named \texttt{batch\_analyze} starts by loading data and models using \textsc{MLFlow}, and runs model inference on the collected data. After that, the pipeline runs the \texttt{run\_evidently} step to check for the presence of data drift or performance degradation and uses the results to trigger a new model training phase. This example shows the benefits of co-using multiple ML frameworks, motivating the analysis we report in the following.



\textbf{MLOps framework co-usage.} As shown in \tabref{tab:usages_patterns}, we found only 29 projects using multiple MLOps frameworks. \textsc{MLFlow} occurs in 11 of the 12 co-usage patterns, while \textsc{Wandb} is never combined with other frameworks. The most common co-usage (\textsc{MLFlow,Prefect}, with 8 occurrences) involves two frameworks that provide different core features. In particular, \textsc{MLFlow} is primarily used for tracking experiments, while \textsc{Prefect} orchestrates the ML pipeline. A similar consideration holds for combining \textsc{MLFlow} with \textsc{Evidently AI}, where experiment tracking is combined with data quality, model monitoring, and validation.
Last but not least, we want to point out that some co-usages combine multiple frameworks, supporting developers in setting up and defining ML pipelines. Looking at these cases, we noticed that \textsc{Kedro} is used to define the ML pipeline steps, while \textsc{Prefect} is mainly used for orchestration.

\subsection{RQ$_3$: MLOps Frameworks Feature Requests}
\figref{fig:taxonomy} depicts the two-level taxonomy of features requested by analyzing the issue reports of the eight MLOps frameworks. Each category in the figure is annotated with two values: the inner number represents the frequency from the LLM-classified dataset, and the outer number is its frequency in the manually-annotated sample. 

The taxonomy consists of \numFirstLevel high-level categories, which are in turn detailed into \numSecondLevel sub-categories. The taxonomy subtrees are also classified by how they map to different MLOps dimensions. These include specific pipeline components (\eg monitoring, deployment), crosscutting concerns (\eg non-functional aspects, infrastructure governance), and other areas such as documentation or user interfaces and experiences. In the following, we will describe each high-level category and their sub-categories.

%
%

\textbf{Experiment Management.} This category includes feature requests/improvements to MLOps runs/experiments, including experiment launching, tagging, creating hierarchies, and accessing metadata.
More specifically, categories are related to: 
\emph{Run lifecycle \& state} (\ie how the experiment execution is managed, \eg paused, restarted, or canceled, and its state observed);  
\emph{Experiment Configuration \& Parameterization}; as well as their \emph{Execution \& Scheduling}, which, among others, also concerns making experiment execution concurrent, or sharing experiment information/results.

\textbf{Metrics and Evaluations.} This category includes requests for the definition or extension of \emph{offline} training and validation metrics, evaluation workflows, and fairness or explainability features. The category excludes runtime monitoring metrics, which belong to a specific category of \emph{Logging and Monitoring}. 
More specifically, sub-categories are related to: \emph{Metric Addition/Customization}, \eg \textsc{deepchecks} issue \#1272 asks for ``Add Cramer's V in addition to PSI for all categorical drift''; 
\emph{Metric lifecycle \& management}, \eg how metrics can be exported or represented; \emph{Metric Clarity improvements}, concerning how to properly convey metrics and make them and their root causes understandable. 

\textbf{CI/CD Pipeline Integration.} This category includes feature requests asking for hooks, steps, or plugins to integrate the framework into existing CI/CD pipelines for automated build, test, and deployment. Sub-categories relate to integrating CI/CD with versioning systems and to CI/CD automation itself. As in issue \#14677 from MLFlow, users ask for the possibility to automatically trigger GitHub workflows from \textsc{MLFlow} events, \ie trigger model deployment when a tag transitions to ``Production''. To achieve this, a webhook needs to be defined in \textsc{MLFlow} and used to run automated testing or deployment. Such a lack may be one of the reasons for the limited use of CI/CD.

\textbf{Dependencies/Environment Management.} This category groups feature requests/enhancements allowing for a proper configuration of the environment in which user code or jobs will execute, and the management of the framework's own software dependencies.

\textbf{API Specification.} This category includes feature requests asking for modifying the way the framework exposes its pieces of functionality through its (Web) APIs or Software Development Kit (SDK). Such requests typically do not involve changing the underlying features but instead modify request/response schemata, parameter and functionality exposure, and hooks. Besides the obvious category related to \emph{API Extensibility} issues, some peculiar requests are related to \emph{Edge Case Handling}, \ie providing custom exceptions or allowing the removal of optional fields, and also \emph{Selective Resource Retrieval}, for adding filtering, pagination, or selective field retrieval to API endpoints that list resources. 

\textbf{Security and Privacy.} This category includes feature requests asking for authentication, authorization, encryption, and secret management handling. 

\textbf{Performance.} This category is about the enhancement of MLOps features---typically without affecting the frameworks' APIs---to improve the efficiency (\eg in terms of reducing latency and improving throughput) in data I/O, resource usage, and in the execution of the MLOps phases. For example, issue \#4350 from \textsc{Kedro} points out the benefits of adding a built-in node caching feature, \ie ``a mechanism to avoid re-executing a node when its inputs, outputs, and logic remain unchanged''.

\textbf{Documentation/Community Management.} This category groups feature requests asking for improving guides, tutorials, API references, code examples, and community engagement channels. It is made up of two sub-categories related to: (i) improving the official \emph{Product Documentation}, \eg API documentation, development guide, contribution guide; and (ii) \emph{Community engagement}, related to how the project interacts with its user community, \eg through external platforms, or by integrating community projects and plugins.

\textbf{Artifact and Data Management.} This category includes feature requests asking for introducing or improving the management of datasets and artifacts \emph{produced by the MLOps framework}, such as models, containers, and runtime reports. 
This includes \emph{Artifact lifecycle/versioning}, related, for example, to how data artifact versions are represented/shown, \eg \textsc{Kedro}'s issue \#1069 asking for linking a ``pipeline output to a version of the pipeline, not just a date/time''; \emph{Artifact usability \& integration}; \emph{Data Catalog Management}, related to how the frameworks' features can access the data catalog; and \emph{Schema Validation}, related to how frameworks enforce schema validation over datasets, \eg `` Increase model signature flexibility to allow for occasionally missing fields while retaining datatype enforcement'' (\textsc{MLFlow} issue \#6783).

\textbf{Third-Party Integration.} This category features requests for new or improved integrations with external services, tools, frameworks, and cloud providers. This includes data connectors, ML framework support, and external API integrations.

\textbf{Model Deployment and Serving.} This category includes requests related to the deployment phase. More specifically, they are related to \emph{Manifest configuration} \ie the creation and management of deployment configuration files; \emph{Deployment Strategy}, \ie detailing how deployment is performed; and  \emph{Deployment Parameters and Triggers Customization} aimed at specifying parameters that control or activate the deployment process and determine its behavior.

\textbf{Model Serialization. } This category features requests for support or improvement of model serialization formats and cross-framework compatibility. This includes handling \emph{Serialization Libraries}; \emph{Serialization Format Support}, \ie supporting specific formats such as  ONNX~\cite{onnx-specification}, TorchScript~\cite{pytorch-jit-doc}, or Pickle~\cite{python-pickle-doc}; and \emph{Model Serialization Format}, \ie the inclusion of specific user-defined information in the model serialization metadata, \eg ``We need to be able to add custom metadata to the `MLmodel` file'' from \textsc{MLFlow} issue \#7235.



\textbf{Logging and Monitoring.} This category relates to 
\emph{runtime observability} of both models, data, and MLOps process status. It includes feature requests about \emph{Add logging information}, related to having information in frameworks' predefined logs, such as injecting custom fields from the runtime context; \emph{Log Customization}, \ie the usual layout, appenders and levels customization, and, above all, a specific category related to \emph{Execution and Data Monitoring}.

\textbf{User Interfaces and Experience.} This category is about requests for improvements to the framework's user interface and experience, for example, to improve the framework's Web or command-line interface. This includes usability, navigation, visual presentation, and how the framework communicates warnings/errors to the user. A peculiar subcategory concerns a uniform representation of the MLOps workflow across different environments.

\section{How feature usages relate to feature requests}
\label{sec:discussion}

This section overviews the mapping between the features used by the MLOps frameworks (RQ$_2$) and those requested by the community (RQ$_3$). Results for RQ$_2$ indicate that users leverage a limited subset of the framework’s features. Therefore, the goal here is to check whether the requests concern unused features (\eg, because they do not work well or are unusable) or whether (as was the case) they request specialization on features already in use. The colors in \figref{fig:taxonomy} represent the mapping between the feature request taxonomy and the key features provided by the frameworks. While some feature requests cannot be directly mapped to features provided by the frameworks, such as Documentation/Community Management or Non-Functional, many requests are not about adding new features but rather about improving the ones already provided. However, we found some exceptions that we will discuss later on in this section. 

For \textsc{Evidently AI}, the feature requests are perfectly aligned with the type of core features it provides. Indeed, users mainly request new data drift detection tests to be added. This justifies co-use with \textsc{Deepchecks}, which focuses more on data integrity checks. For \textsc{BentoML}, most requests relate to the deployment phase. As confirmed by the usage analysis, developers mainly use the framework to streamline model serving and deployment, and they request enhancements to ensure a ``secure deployment'' that minimizes information disclosure (issue \#4663). 

\textsc{Prefect}, is mainly used to set up and define the ML pipeline, and for workflow orchestration.
Feature requests indicate that users are often dissatisfied with the in-place governance and observability mechanisms. For example, issue \#3484 asks for allowing the definition of ``resilient and cost-effective" pipelines, in which it is possible to stop the execution of a specified flow in the presence of failed tasks, rather than waiting for all tasks in the flow to complete. This is closely related to the framework's monitoring feature, which, as explained in RQ$_2$, is better suited for pipeline health monitoring than for live monitoring in production.
 In this regard, issue \#884 asks for allowing users to add \textit{runtime logs} to tasks created with the \texttt{@task} decorator.
On the same line,  \textsc{Kedro} is not used for monitoring. However, we found issues, like \#628 (``I would like to migrate the code to use Kedro. One thing that is important for us is to be able to track the progress so that we can provide that info to the end user.  I think being able to track the progress of nodes or pipelines would be a really nice feature for Kedro.''), where users request the possibility to track the progress of nodes or pipelines, \ie monitoring of the pipeline health.

As found in RQ$_1$, the study's frameworks are mainly used by invoking their APIs in a Python script. This aligns with the high number of feature requests belonging to this category. As an example, the latest version of \textsc{MLFlow} allows developers to directly tag a current active experiment by using \texttt{mlflow.set\_experiment\_tag()}. In older versions, developers needed to manually instantiate an \texttt{MLFlowClient()} to achieve the same goal. Issue \#6083 requested moving this feature into the fluent API, mainly because setting tags for experiments is a key operation for frameworks supporting ML experiment tracking. Also, there are requests to better document frequent usage patterns of the framework.
Often, developers ask for code examples and detailed documentation for each exposed API. Concerning the latter, issue \#1565 from \textsc{Deepchecks} emphasizes the struggle of users who want to set up test suites for validating their models, but, unfortunately, this information is missing, \ie ``User should know how to run all built-in suites from docstrings, and should know what it contains without needing to run it''. 

To summarize, the previously discussed examples highlight the barriers/challenges developers encounter when integrating and using MLOps frameworks. This may justify the low adoption rate observed in Python open-source projects. Of course, this topic deserves a further, fine-grained investigation.

\section{Threats to Validity}
\label{sec:threats}

Threats to \emph{construct validity} concern the relationship between theory and observation. They may be primarily related to how we capture information about framework usage and feature requests. In principle, developers could also use frameworks locally, while we mainly capture the use from scripts (including GitHub Workflows) and programs in the repository. However, based on what we discovered, the main use of MLOps frameworks is within the systems themselves.
Regarding feature requests, it is known that labels in issue trackers may not be accurate \cite{AntoniolAPKG08}. While we may miss some feature requests, we avoid false positives through the detailed analysis done on the collected issues. 

Threats to \emph{internal validity} concern factors internal to our study that could have influenced our results. The identification of dependents can be influenced by dependency graph accuracy \cite{BifolcoRNFSP25}. However, in the worst case, this may have introduced some false negatives rather than false positives, given the following code analysis, which checked for the actual MLOps framework usage. 
The usage analysis could be affected by code analysis errors or approximations. However, we randomly selected some of the analyzed clients and double-checked the automatically retrieved usages against those derived manually.
Regarding the manual categorization of APIs (RQ$_2$), we acknowledge that it may still have led to mistakes despite the double-check.
The taxonomy for feature requests (RQ$_3$) has been created by combining a manual analysis with a follow-up LLM-based automated classification, as previous work did \cite{AhmedDTP25}. We have assessed the reliability of our coding using Cohen's Kappa inter-rater agreement.

Threats to \emph{external validity} concern the generalizability of our findings. Our study focuses on eight popular open-source MLOps frameworks, selected based on the criteria described in \secref{sec:mlopstools}. While being diverse in terms of provided features, such frameworks do not represent the universe of open-source frameworks and, for sure, they do not cover commercial tools. Also, for observability reasons, the study limits the analysis to open source clients of the MLOps frameworks. These, again, may not reflect the maturity of many closed-source ML-intensive applications.

\section{Related Work}
\label{sec:related}

This section discusses literature about characterizing MLOps processes and studies on MLOps usage in the wild. A broader analysis of MLOps challenges can be found in extensive literature reviews \cite{3747346,10.1145/3625289}.

\subsection{Characterizing MLOps processes}
The work most closely related to ours is the one from Calefato \etal~\cite{calefato}, who investigate how (and how much) open-source GitHub projects actually adopt MLOps practices, focusing particularly on two automation tools, namely, GitHub Actions (the built-in CI/CD system in GitHub) and CML (Continuous Machine Learning)---a specialized framework that extends GitHub Actions for ML-specific tasks---reaching a resounding conclusion: the adoption of MLOps at scale is limited. This conclusion provides strong motivation behind our work, which instead aims to go one step further by characterizing open-source MLOps frameworks to encourage their effective use in context. From a similar perspective, Warnett \etal~\cite{warnett} aim to strengthen MLOps adoption and complexity management by investigating whether supplementing informal descriptions of MLOps system architectures with semi-formal UML-based diagrams improves participants’ understanding and use of those systems. In the same way, we aim to enhance the user experience with MLOps frameworks by further characterizing them, drawing on open-source quantitative data, while Warnett \etal~\cite{warnett} conduct controlled experiments featuring semi-formal MLOps descriptions. The work and its key result---describing MLOps frameworks more formally aids their adoption in context---further underscores the need for greater characterization and descriptive capacities for MLOps to support their practical use. Along the same lines, John \etal~\cite{john} characterize MLOps frameworks and automations with a descriptive framework and its prescriptive maturity model, both designed drawing from a systematic literature review (SLR) and validated over three industrial case-studies, while Recupito \etal~\cite{recupito} carry out a similar work with the intent of offering a comparative analysis. Overall, these papers also intend to characterize MLOps, with a key difference from our scope, that is, their (limited) industrial validation focus versus our stronger open-source characterization target.

\subsection{Empirical Studies on MLOps in the Wild}

Makinen \etal~\cite{makinen} evaluate the features and facilities that MLOps systems offer from an organizational and technical perspective, providing an overview of how to arrange and embed available tools and practices, drawing on a large-scale survey. Kumara \etal~\cite{kumara} and Matsui and Goya~\cite{matsui} offer a practical overview of MLOps adoption and its application in action. Their work can be used to bootstrap MLOps adoption in practice, while ours focuses more on refining the organisational and technical decisions around which framework to use and for what purpose. 
Last but not least, Zhao \etal~\cite{ZhaoCBAH24} automatically analyzed using BERTtopic~\cite{grootendorst2022bertopic} over 15k Stack Overflow posts and issues on GitHub/GitLab to study discussions related to the management of ML assets, classifying problem and solution inquiries. While we share the analysis of issues, the purpose is different, as our goal is to identify desired features and improvements. Also, we manually elicited the taxonomy of requested features, and only leveraged the LLMs for a paired, second round of classification.

\section{Conclusion and Work-in-Progress}
\label{sec:conclusion}

This paper investigated the use of eight popular open-source MLOps frameworks in projects hosted on GitHub. By analyzing, on the one hand, the framework usages by dependent projects and, on the other hand, their feature requests, we found that:
\begin{compactitem}
\item \textbf{RQ$_1$:} Frameworks are mainly used as APIs rather than from their CLI or GitHub workflows. Furthermore, frameworks for model tracking and for governing the whole ML pipelines are the most widely used.
\item \textbf{RQ$_2$:} Framework usages cover different phases, including model training, monitoring, and infrastructure governance. At the same time, developers tend to combine multiple frameworks to complement the different features they provide.
\item \textbf{RQ$_3$:} We identified a taxonomy of \numFirstLevel feature request categories, in turn detailed into \numSecondLevel sub-categories, covering, on the one hand, aspects throughout an MLOps pipeline and, on the other hand, a better exposure of APIs, of their documentation, as well as a proper CLI or CI/CD support.
\item \textbf{Mapping between feature usage and requests:} Feature requests are generally aimed at improving the main, specific type of feature a framework was conceived for, rather than introducing completely different ones.
\end{compactitem}

The study results have implications for different stakeholders. \textbf{Practitioners} can contrast their usage patterns with those from the wild, \eg understanding how multiple tools are combined to cover different MLOps aspects. \textbf{Tool creators} can better understand how to design their framework, considering typical change direction avenues and, possibly, interoperability requirements with other tools, API exposure, and availability as CLI or CI/CD tools. \textbf{Researchers} could develop recommenders to aid practitioners in configuring and (co-) using MLOps frameworks.

Our work-in-progress aims to complement this open-source study by conducting analyses in industrial environments through interviews and surveys. 

\section{Data Availability}
The study datasets and scripts are available in our replication package \cite{replication}.

\section*{Acknowledgments}
Di Penta acknowledges the Italian ‘‘PRIN 2022’’ project TRex-SE: ‘‘Trustworthy Recommenders for Software Engineers’’, grant n. 2022LKJWHC. 

\balance
\bibliographystyle{ACM-Reference-Format}
\bibliography{main}

\end{document}